\documentclass[aps,prl,twocolumn,showpacs,floatfix,superscriptaddress]{revtex4-1}
\usepackage{graphicx,epsfig,color,dcolumn}

\begin{document}

\title{Transport properties of molecular junctions from many-body perturbation theory}

\author{T. \surname{Rangel}}
\email[]{tonatiuh.rangel@uclouvain.be}
\affiliation{European Theoretical Spectroscopy Facility (ETSF)}
\affiliation{Institute of Condensed Matter and Nanosciences (IMCN), Universit\'e Catholique de Louvain, Place Croix du Sud 1 bte 3, B-1348 Louvain-la-Neuve Belgium}
\author{A. Ferretti}
\affiliation{Department of Materials, University of Oxford, Oxford, United Kingdom}
\affiliation{CNR-INFM S3 and Dipartimento di Fisica, Universit\`a di Modena e Reggio Emilia, Modena, Italy}
\affiliation{European Theoretical Spectroscopy Facility (ETSF)}
\author{P. E. Trevisanutto}
\affiliation{European Theoretical Spectroscopy Facility (ETSF)}
\affiliation{Max-Planck Institut f\"ur Mikrostrukturphysik, Halle/Saale, Germany}
\affiliation{Institut N\'eel, CNRS and UJF, Grenoble, France}
\author{V. Olevano}
\affiliation{European Theoretical Spectroscopy Facility (ETSF)}
\affiliation{Institut N\'eel, CNRS and UJF, Grenoble, France}
\author{G.-M. Rignanese}
\affiliation{European Theoretical Spectroscopy Facility (ETSF)}
\affiliation{Institute of Condensed Matter and Nanosciences (IMCN), Universit\'e Catholique de Louvain, Place Croix du Sud 1 bte 3, B-1348 Louvain-la-Neuve Belgium}

\date{\today}

\begin{abstract}

The conductance of single molecule junctions is calculated using a Landauer approach combined to many-body perturbation theory (MBPT) to account for electron correlation.
The mere correction of the density-functional theory eigenvalues, which is the standard procedure for quasiparticle calculations within MBPT, is found not to affect noticeably the zero-bias conductance. 
To reduce it and so improve the agreement with the experiments, the wavefunctions also need to be updated by including the non-diagonal elements of the self-energy operator.

\end{abstract}


\maketitle

Most recent theoretical studies of coherent transport in nanojuctions are based on a Landauer approach~\cite{Datta}, in which the electron interactions are treated at a simplified mean-field level using density-functional theory (DFT).
While this approach has proven quite successful for systems having a strong coupling between the molecule and the metallic leads~\cite{thygesen.2005,garcia-suarez.2005}, it overestimates the zero-bias conductance of weakly coupled systems (by up to 3 orders of magnitude) compared to experimental measurements~\cite{diventra.prl.2000,quek.nano.2007,nitzan.2003}.
Many explanations have been proposed for such a discrepancy. For example, arguing an uncertainty over the experimental junction structure, the sensitiveness to the contact geometry was investigated~\cite{emberly.2001,ning.2010}.

More fundamentally, it has been found that the zero-bias conductances may vary by several orders of magnitude when using different exchange-correlation (XC) energy functionals~\cite{ke.2007,toher.2008}.
Moreover, the DFT lack of derivative discontinuity was shown to be a source of significant errors in weakly coupled systems~\cite{koentopp.2006}.
In fact, the validity of using DFT to treat electron interactions in the Landauer formalism has also been questioned~\cite{mera.2010}, and it is believed that a many-body theory should be better suited.
Indeed, the DFT eigenstates do not have a formal direct connection to the quasiparticle (QP) states.

Recently, many-body perturbation theory (MBPT) calculations~\cite{hedin.1965} have been used in combination with the Landauer formalism to study the conductance of a simple gold monoatomic chain~\cite{darancet.2007}.
The applicability of this method to more complex systems, as realistic three dimensional junctions, is still out of reach.
Therefore, a simpler approach relying on a model self-energy~\cite{quek.nano.2007,mowbray.2008}, based on projectors onto molecular orbitals, has been applied to molecular junctions with some success.
Nevertheless, a formal justification of this model is still lacking.

In this letter, the effect of electron-electron interactions on the zero-bias conductance is investigated for the benzene diamine (BDA) and benzene dithiol (BDT)
molecules attached to gold electrodes.
The true QP electronic structure, calculated using different approximations for the MBPT self-energy, is used in the Landauer formalism.
It is found that correcting only the eigenenergies has no or little effect on the conductance.
To improve the agreement with the experiment, it is thus crucial to update also the wavefunctions by taking into account the off-diagonal self-energy matrix elements.
A comparison of the original and the updated local DOS (LDOS) at the Fermi level shows that both the molecule and the gold regions are affected.
In particular, it is found that the change of the zero-bias conductance is triggered by a decrease of the molecular character and an increase of the  $e_g (d_{z^2})$ character of the wavefunctions on the gold atoms.
Finally, our results are compared to those provided by the molecular-projectors model proposed in Refs.~\cite{quek.nano.2007,mowbray.2008}.

The DFT and MBPT calculations are performed using the \textsc{Abinit} package~\cite{gonze.2009}.
The XC energy is approximated by the PBE functional~\cite{perdew.1996}.
Norm-conserving pseudopotentials~\cite{troullier.1991} are used.
For gold, these include the 5$s$ and 5$p$ semicore states which are crucial for the MBPT calculations~\cite{marini.2001}. 
The wavefunctions are expanded on a plane-wave basis set up to kinetic energy cutoff of 30 Ha.
The corresponding maximally-localized wannier functions (MLWFs)~\cite{mlwfs} are obtained following the procedure of Ref.~\cite{hamann.2009}.
These are used in the transport calculations which are performed with the \textsc{WanT} package~\cite{want}. 

Two systems are considered: BDA and BDT attached to gold electrodes~\cite{note:bdt_geometry}.
For the atomic relaxation, a 2$\times$2 surface cell is adopted for the Au (111) surface with seven atomic layers in the electrodes.
The relaxed geometries agree with previous theoretical predictions~\cite{mowbray.2008,ning.2007,ning.2010}.
For the MBPT calculations, three gold layers are removed.
This guarantees an affordable computational cost without compromising the calculated conductance~\cite{note:cell_size} despite the rather packed geometry (with $\sim$3.1~\AA~between the repeated images of the molecules). 
All the parameters of the calculations are converged to ensure an error smaller than 0.001~$\mathcal{G}_0$ on the zero-bias conductance $\mathcal{G}$($E$=0).
Hence, a 8$\times$8$\times$3 grid of $\mathbf{k}$ points is adopted to sample the Brillouin zone.
The QP corrections are calculated explicitly for 210 bands at 96 irreducible $\mathbf{k}$-points including $\sim$300 bands in the calculations. 

For the MBPT calculations, the $GW$ approximation is adopted for the self-energy $\Sigma$, neglecting vertex corrections~\cite{hedin.1965}.
In principle, $\Sigma$ is a non-hermitean, non-local and frequency-dependent operator.
The non-hermitian part of $\Sigma$ is ignored, assuming infinite QP lifetimes; while the non-locality is fully taken into account.
Different flavors are then considered for the $GW$ self-energy.
In the standard single-shot $G_0W_0$ approach, the self-energy is approximated using the DFT electronic structure; and, the QP corrections to the DFT eigenvalues are determined using first-order perturbation theory considering only the diagonal elements of $\Sigma$ in the space of the DFT orbitals  $\phi^{\rm sys}_i$ of the contacted-molecule system.
The frequency dependence is obtained using a plasmon pole model~\cite{godby.1989}. 
In order to calculate the first order correction to the DFT wavefunctions, the off-diagonal elements of $\Sigma$ would also be needed.
Currently, such calculations are out of reach for the systems considered here.
Nevertheless, a full diagonalization can be performed if a further simplified self-energy is considered.
For this purpose, the Coulomb-hole screened-exchange approach~\cite{hybertsen.1986}, which is a static approximation to the $GW$ self-energy, is adopted here.
For the sake of comparison, both the full ($\overline{\textrm{CHSX}}$) and diagonal (CHSX) forms of this self-energy are considered.
Finally, the molecular-projectors model ($\overline{\textrm{MPM}}$) proposed in Refs.~\cite{quek.nano.2007,mowbray.2008} is also used for the self-energy:
\begin{equation}
\Sigma(E) = \sum_m {\rm sgn}(E-E_F) \Delta | \phi^{\rm mol}_m \rangle \langle
\phi^{\rm mol}_m | ,
\end{equation}
where $m$ runs over the DFT orbitals $\phi^{\rm mol}_m$ of the isolated molecule, $E_F$ is the Fermi energy, and $\Delta$ is the scissor-operator shift to be applied.
Its value is obtained in a two-step procedure.
First, the correction to the DFT HOMO-LUMO gap of the molecule in the gas phase is computed either from a $\Delta$SCF calculation or, more simply, by matching the experimental value. 
Second, the presence of the metallic surface is taken into account through a classical image-charge model leading to a reduction of $\Delta$ compared to the gas phase, as calculated using the $G_0W_0$ approximation~\cite{neaton.2006,lastra.2009}.
This self-energy operator is non-diagonal in the space of the DFT orbitals  $\phi^{\rm sys}_i$ of the contacted-molecule system.
For the sake of comparison, its diagonal form (MPM) is also considered.

\begin{figure}[h]
\includegraphics{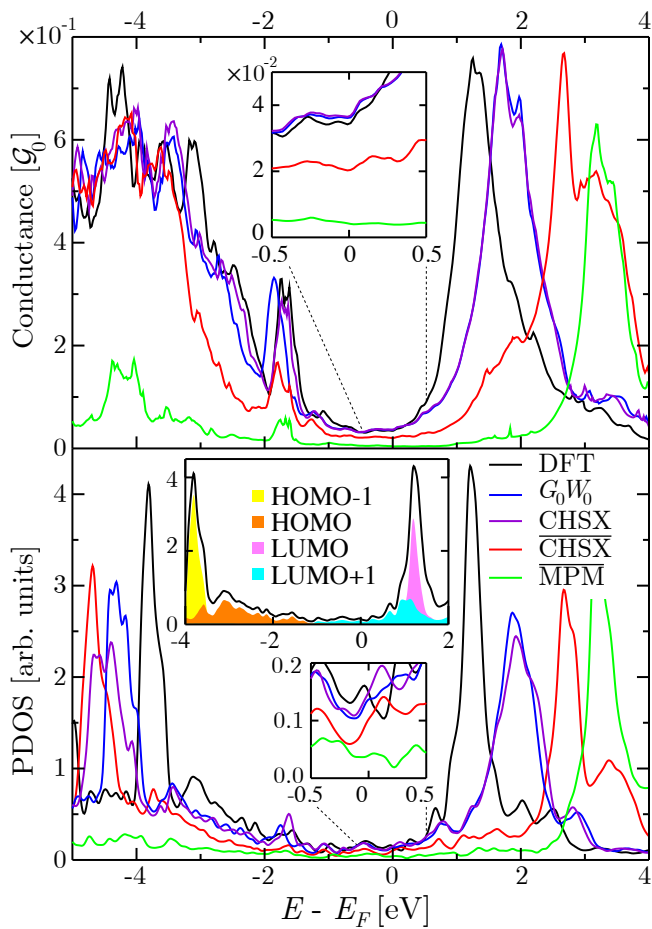}
\caption{Conductance (top) and PDOS on the molecule (bottom) as a function of the energy (the Fermi energy $E_F$ is set to zero) for the BDT junction.
Different approaches are considered: DFT (in black), $G_0W_0$ (in blue), CHSX (in purple), $\overline{\textrm{CHSX}}$ (in red), and $\overline{\textrm{MPM}}$ (in green).
For each panel, an inset provides a zoom around the Fermi energy.
In the bottom panel, a second inset shows the total PDOS on the molecule computed in DFT (black line),
and the corresponding contributions of the HOMO-1 (in yellow), the HOMO (in orange), the LUMO (in pink), and the LUMO+1 (in cyan).}
\label{fig:cond}
\end{figure}

The values of the zero-bias conductance $\mathcal{G}$($E$=0) calculated with these different approaches are reported in Table~\ref{table.cond} for BDA and BDT.
For the latter, the complete energy dependence of the conductance $\mathcal{G}$($E$) is also given in Fig.~\ref{fig:cond}.
The DFT results are in good agreement with previous calculations~\cite{diventra.prl.2000,quek.nano.2007,nitzan.2003,ning.2007,ning.2010}.
Quite surprisingly, the $G_0W_0$ results for $\mathcal{G}$($E$=0) are almost identical to the DFT ones despite the changes in the curve $\mathcal{G}$($E$).

\begin{table}[h]
\begin{ruledtabular}
\begin{tabular}{lcccccccc}
       &DFT &$G_0W_0$ &CHSX &$\overline{\textrm{CHSX}}$ &MPM  &$\overline{\textrm{MPM}}$ & Expt.\\
\hline
BDA & 0.018 & 0.019 & 0.019 & 0.013 & 0.017 & 0.004
&0.006\tablenote{References~\onlinecite{quek.nano.2007,venkataraman.2006}.}
\\
BDT & 0.034 & 0.036 & 0.037 & 0.020 & 0.027 & 0.004
&0.011\tablenote{References~\onlinecite{xiao.2004,tsutsui.2006}.}
\end{tabular}
\end{ruledtabular}
\caption{Zero-bias conductance $\mathcal{G}$($E$=0) in $\mathcal{G}_0$ units calculated with different approaches (see text) for the two systems.}\label{table.cond}
\end{table}

In order to understand this finding, the projected density of states (PDOS) on the molecule is computed as reported in the bottom panel of Fig.~\ref{fig:cond} for BDT.
In particular, an inset shows a decomposition of the total PDOS on the molecule calculated within DFT in terms of the contributions of the different molecular orbitals close to the Fermi level.
The HOMO-1 and LUMO (in yellow and pink, respectively) lead to very sharp peaks in the PDOS.
In fact, these states are very localized on the molecule and present very little hybridization with the leads.
Hence, they do not play an important role on the conductance at 0 eV.
The latter is thus mainly driven by the HOMO and LUMO+1 (in orange and cyan, respectively) in agreement with the findings of Ref.~\cite{ning.2007}.

As expected, the $G_0W_0$ corrections open the gap between the occupied and unoccupied molecular-like levels, as can be seen in the PDOS (bottom panel) and the conductance (top panel) in Fig.~\ref{fig:cond}.
This opening of the gap is much smaller than the one obtained for the isolated molecule ($\sim$5.5 eV).
In particular, the gap between the HOMO and LUMO+1 is increased only by $\sim$0.5~eV.
The presence of the metallic surface can be estimated to account for a reduction of $\sim$2 eV with respect to the isolated molecule~\cite{neaton.2006,lastra.2009}.
More importantly, the $G_0W_0$ corrections depend on the weights of the different states of the system on the molecular orbitals $w_i= |\langle \phi^{\rm sys}_i | \phi^{\rm mol}_m \rangle|^2$.
These are quite small for the very hybridized HOMO and LUMO+1, hence the small gap opening.
In contrast, these weights are higher for the more localized LUMO and HOMO-1.
Thus, the correction to their gap ($\sim$1.2~eV) is more than twice bigger than for the HOMO and LUMO+1.
Similar results are observed for BDA.

Due to this very limited opening of the gap, the resulting $G_0W_0$ zero-bias conductance $\mathcal{G}$($E$=0) is almost equal to DFT one for both systems.
A similar result is observed when using the CHSX approximation, even though this approximation usually leads to bigger gaps than those obtained with the $G_0W_0$ approach.
In conclusion, the conductance $\mathcal{G}$($E$=0) is almost not affected compared to DFT when the self-energy only corrects the eigenvalues. 

When updating not only the eigenenergies but also the wavefunctions by the $\overline{\textrm{CHSX}}$ approximation, $\mathcal{G}$($E$=0) is reduced by a factor 1.4 (1.7) in the direction of the experimental value reaching 0.013 (0.020) $\mathcal{G}_0$ for BDA (BDT).
The agreement with experiments would probably be further improved if one could afford to perform calculations with 
(i) more than one step in the self-consistency loop on the wavefunctions; 
(ii) more realistic geometries, such as those depicted in Ref.~\cite{quek.nano.2007}; and
(iii) a higher number of gold layers entering the Landauer formula.

\begin{figure}
\includegraphics{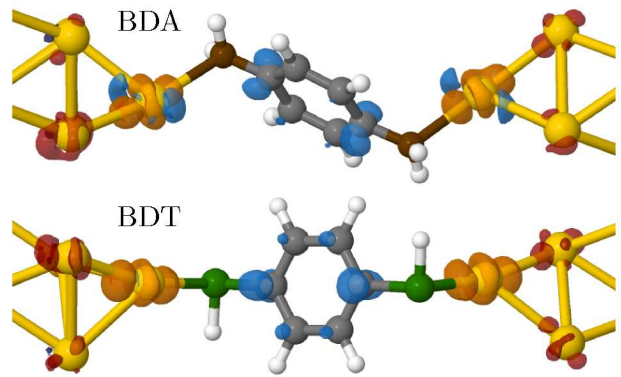}
\caption{Difference between the $\overline{\textrm{CHSX}}$ and CHSX LDOS calculated in an energy window of 0.8~eV around the Fermi level for BDA (top) and BDT (bottom).
Four isovalues are represented: $+4\rho$ in orange, $+1\rho$ in dark red, $-1\rho$ in dark blue, and $-4\rho$ in light blue, with $\rho=4\times10^{-4}$ ($6\times10^{-4}$) $e^{\textrm{-}}$/\AA$^3$ in BDA (BDT).
The Au, C, N, S, and H atoms are represented by yellow, gray, brown, green, and white spheres, respectively.
}
\label{fig:ldos}
\end{figure}

The PDOS on the molecule (Fig.~\ref{fig:cond}) shows that $\overline{\textrm{CHSX}}$ corrections contribute to further separate the unoccupied and occupied molecular-like orbitals.
Since the eigenenergies are only affected at the second order compared to CHSX, this increased opening of the gap is to be attributed to modifications of the metal-molecule hybridizations.
To gain more insight on the changes of the wavefunctions, the LDOS is computed in an energy window of 0.8~eV around the Fermi level $E_F$:
\begin{equation}
{\rm LDOS}(\mathbf{r})= \int_{E_F - 0.4 eV}^{E_F + 0.4 eV} \sum_i |\phi^{\rm sys}_i(\mathbf{r})|^2 \delta(E-E_i) dE.
\end{equation}
The effect of the off-diagonal elements of $\Sigma$ can be analyzed by plotting the difference between the $\overline{\textrm{CHSX}}$ and CHSX LDOS, as shown in Fig.~\ref{fig:ldos}.
Two main changes can be identified.
First, {\it the molecular character is reduced}.
This charge transfer does not take place in space (it is less 0.02$e^-$ from the molecule to the gold leads) but in energy:
the molecular orbitals being shifted away from $E_F$ (see the PDOS in Fig.~\ref{fig:cond}).
Second, {\it the $e_g (d_{z^2})$ character increases on gold atoms} (see the typical ring shaped red lobes on the gold atoms).
For the Au add-atoms, the $z$ direction of the $e_g (d_{z^2})$ orbital is oriented along the Au-N bonding direction.
For the atoms of the next gold layer, it changes though it is still conditioned by the Au-N direction.
Since there is not a noticeable reduction of the gold character in other directions and no molecule-gold charge transfer, this increase of the $e_g (d_{z^2})$ character can be mainly to a transfer of electrons from lower energies to the $E_F$ region due to the complete diagonalization of the self-energy corrected hamiltonian.
In contrast, the difference between the DFT and CHSX LDOS (not shown here) is negligible on the molecule, which explains their similar zero-bias conductance. 

Finally, the self-energy is also modeled by the $\overline{\textrm{MPM}}$~\cite{quek.nano.2007,mowbray.2008}, which includes off-diagonal elements, using $\Delta$=2~eV.
The $\overline{\textrm{MPM}}$ leads to an effective opening of the gap between the unoccupied and occupied molecular-like orbitals which is much bigger ($\sim$4~eV) than the one obtained with the $\overline{\textrm{CHSX}}$ approximation ($\sim$2~eV).
As a result, the $\overline{\textrm{MPM}}$ conductance is much lower than in any of the fully {\it ab initio} approaches.
This discrepancy can be attributed to some limitations in both the fully {\it ab initio} calculations and the model ones.
On the one hand, the use of highly-packed geometries considerably increases the screening and reduces the size of the corrections obtained within the fully {\it ab initio} calculations.
Nevertheless, the use of the $\overline{\textrm{CHSX}}$ approximation, which is known to overestimate the experimental gaps, may partially compensates this reduction.
On the other hand,  two main drawbacks can be pointed out for the model calculations:
(i) the classical image charge model is not always appropriate in predicting the individual shifts~\cite{lastra.2009};
(ii) many-body effects on the gold region as observed in our calculations within the $\overline{\textrm{CHSX}}$ approximation. 

In summary, we have demonstrated that a self-energy operator leading to the mere correction of the eigenenergies (such as the $G_0W_0$ or CHSX approximations) is not enough to change the initial DFT zero-bias conductance.
In contrast, when updating also the wavefunctions (such as obtained by the $\overline{\textrm{CHSX}}$ approximation),  the conductance is reduced, improving the agreement with the experiment.
The conductance change can be attributed to both a reduction of the molecular character and an increase of the $e_g  (d_{z^2})$ gold character.
Finally, the disagreement between the $\overline{\textrm{MPM}}$ and {\it ab initio} calculations originates from the value of the scissor-operator shift $\Delta$ and from the wavefunction changes on the gold atoms.

\begin{acknowledgments}
This work was supported by the EU FP6 and FP7 through the Nanoquanta NoE (NMP4-CT-2004-50019) and the ETSF I3 e-Infrastructure (Grant Agreement 211956), and the project FRFC N$^\circ$. 2.4502.05.
We also thank the the French Community of Belgium for financial support via the Concerted Research Action programme (ARC NANHYMO: convention 07/12-003).
\end{acknowledgments}


\begin{thebibliography}{98}
%

\bibitem{Datta}
S.~Datta,
{\it Electronic Transport in Mesoscopic Systems},
Cambridge University Press, New York, 1995.

\bibitem{thygesen.2005}
K. S. Thygesen, K. W. Jacobsen,
Phys. Rev. Lett. {\bf 94}, 036807 (2005).

\bibitem{garcia-suarez.2005}
V. M. Garc\'ia-Su\'arez, A. R. Rocha, S. W. Bailey, C. J. Lambert, S. Sanvito,  and J. Ferrer,
Phys. Rev. Lett. {\bf 95}, 256804 (2005).

\bibitem{diventra.prl.2000}
M. Di Ventra, S. T. Pantelides, and N. D. Lang,  
Phys. Rev. Lett. {\bf 2000}, 84, 979.

\bibitem{quek.nano.2007}
S. Y. Quek, L. Venkataraman, H. J. Choi, S. G. Louie, M. S. Hybertsen and J. B. Neaton,
Nano Letters {\bf 7}, 3477 (2007).

\bibitem{nitzan.2003}
A. Nitzan and M. A. Ratner,
Science {\bf 300},1384 (2003); and references therein.

\bibitem{emberly.2001}
E. G. Emberly and G. Kirczenow,
Phys. Rev. B, {\bf 64}, 235412 (2001).

\bibitem{ning.2010}
Z. Ning, W. Ji, and H. Guo,
arXiv:0907.4674v2.

\bibitem{ke.2007}
S.H. Ke, H.U. Baranger, and W. Yang,
J. Chem. Phys. {\bf 126}, 201102 (2007).

\bibitem{toher.2008}
C. Toher and S. Sanvito,
Phys. Rev. B {\bf 77}, 155402 (2008).

\bibitem{koentopp.2006}
M. Koentopp, K. Burke and F. Evers, 
Phys. Rev. B {\bf 73}, 121403(R) (2006).

\bibitem{mera.2010}
H. Mera, Y.M. Niquet, 
Phys. Rev. Lett. {\bf 105}, 216408 (2010).

\bibitem{hedin.1965}
L. Hedin,
Phys. Rev. {\bf 139}, 796 (1965).

\bibitem{darancet.2007}
P. Darancet, A. Ferretti, D. Mayou, and V. Olevano,
Phys. Rev. B \textbf{75}, 075102 (2007).

\bibitem{mowbray.2008}
D. J. Mowbray, G. Jones, and K. S. Thygesen,
J. Chem. Phys. {\bf 128}, 111103 (2008).

\bibitem{gonze.2009}
X. Gonze {\it et al.},
Computer Physics Communications, {\bf 180}, 2582 (2009).

\bibitem{perdew.1996}
 J. P. Perdew, K.Burke, M.Ernzerhof,
Phys. Rev. Lett. {\bf 77}, 3865 (1996).

\bibitem{troullier.1991}
N. Troullier and J. L. Martins,
Phys. Rev. B {\bf 43}, 1993 (1991).

\bibitem{marini.2001}
A. Marini, G. Onida and R. Del Sole,
Phys. Rev. Lett., {\bf 88}, 016403 ( 2001).

\bibitem{mlwfs}
I. Souza, N. Marzari and D. Vanderbilt,
Phys. Rev. B {\bf 65}, 035109 (2001);
N. Marzari and D. Vanderbilt,
Phys. Rev. B {\bf 56}, 12847 (1997).

\bibitem{hamann.2009}
D. R. Hamann and D. Vanderbilt,
Phys. Rev. B {\bf 79}, 045109 (2009).

\bibitem{want}
WanT code by A. Ferretti, B. Bonferroni, A. Calzolari, and M. Buongiorno Nardelli, (http://www.wannier-transport.org).

\bibitem{note:bdt_geometry}
For BDT, we consider a system in which the H atoms remain attached to the thiols after the molecule has being absorbed to the Au surfaces. 

\bibitem{ning.2007}
J. Ning, R. Li, X. Shen, Z. Qian, S. Hou, A. R. Rocha and S. Sanvito,
Nanotechnology {\bf 18}, 345203 (2007).

\bibitem{note:cell_size}
For the BDT geometry, it was explicitly checked that using a 3$\times$3 surface cell and 7 layers of gold does not change significantly the transport properties at the DFT and $G_0W_0$ levels.
Note also that, for both systems, the coupling self-energies (between the system to ideal infinite leads) are calculated at the DFT level in order to limit the effect of some errors introduced by the use of only four gold layers.

\bibitem{godby.1989}
R.W. Godby and R.J. Needs,
Phys. Rev. Lett. {\bf 62}, 1169, (1989).

\bibitem{hybertsen.1986}
M. S. Hybertsen and S. G. Louie,
Phys. Rev. B {\bf 34}, 5390 (1986).

\bibitem{neaton.2006}
J. B. Neaton, Mark S. Hybertsen and S. G.  Louie,
Phys. Rev. Lett., {\bf 97}, 216405 (2006).

\bibitem{lastra.2009}
 J. M. Garcia-Lastra, C. Rostgaard, A. Rubio and K. S. Thygesen,
 Phys. Rev. B {\bf 80}, 245427 (2009).

\bibitem{venkataraman.2006}
L. Venkataraman, J. E. Klare, I. W. Tam, C. Nuckolls, M. S. Hybertsen, M. L. Steigerwald,
Nano Letters, {\bf 6}, 458 {2006};
L. Venkataraman, J. E. Klare, C. Nuckolls, M. W. Hybertsen and M. L. Steigerwald,
Nature {\bf 442}, 904 (2006).

\bibitem{xiao.2004}
X. Xiao, B. Xu, and N. J. Tao,
Nano Lett. {\bf 4}, 267 (2004).

\bibitem{tsutsui.2006}
M. Tsutsui, M. Taniguchi and T. Kawai,
Nano Lett. {\bf 9}, 2433 (2009);
M. Tsutsui, Y. Teramae, S. Kurokawa and A. Sakai,
Appl. Phys. Lett. {\bf 89}, 163111 (2006).

\end{thebibliography}

\end{document}